\documentclass{ws-ijmpa}

\begin{document}

\markboth{L. R. Dai, Z. Y. Zhang, Y. W. Yu} {The Extended Chiral
SU(3) Quark Model}

%
\catchline{}{}{}{}{}
%

\title{ The Short Range Mechanism of N-N interaction \\
 in the Extended Chiral SU(3) Quark
Model \footnote{The project is supported by Scientific Research
Foundation of Liaoning Education Department (No. 202122028)}}

\author{\footnotesize L. R. Dai\footnote{dailr@lnnu.edu.cn}}

\address{Department of Physics, Liaoning Normal University,
116029, Dalian, P. R. China}

\author{Z. Y. Zhang, Y. W. Yu}

\address{Institute of High Energy Physics, 100039, Beijing,
P. R. China }

\maketitle


\begin{abstract}
We give the comparisons between the chiral $SU(3)$ quark model and
the extended chiral SU(3) quark model. The results show that the
phase shifts of NN scattering are very similar. However, the short
range mechanisms of nucleon-nucleon interaction are totally
different. In the chiral $SU(3)$ quark model, the short range
interaction is dominantly from OGE, and in the extended chiral
$SU(3)$ quark model, it is dominantly from vector meson exchanges.
\end{abstract}

\section{Introduction}

   The chiral $SU(3)$ quark model \cite{Zhang} was proposed by
generalizing the idea of the $SU(2)$ $\sigma$ model to the flavor
$SU(3)$ case. In this model, the nonet pseuo-scalar meson
exchanges and the nonet scalar meson exchanges are considered in
describing the medium and long range parts of the interactions,
and the one gluon exchange (OGE) potential is still retained to
contribute the short range repulsion. By using this model, the
energies of the baryon ground states, the N-N scattering phase
shifts and the hyperon-nucleon (Y-N) cross sections can be
reproduced reasonably. It seems that the repulsive core of the N-N
interaction can be explained by the OGE and the quark exchange
effect.

Since last few years, Shen et al \cite{Shen}, Riska and Glozman
\cite{Glozman} applied the quark-chiral field coupling  model to
study the baryon structure. They found that the chiral field
coupling is also important in explaining the structure of baryons.
In the work of Riska et al \cite{Glozman}, the vector meson
coupling was also included to replace OGE. They pointed out the
spin-flavor interaction is important in explaining the energy of
the Roper resonance and got a comparatively good fit to the baryon
spectra. The quark-chiral field coupling model  is  a big
challenge to the Isgur's model \cite{Isgur}, in which the OGE
governs the baryon structure.

On the other hand, in the traditional one boson exchange (OBE)
model on baryon level,the N-N short range repulsion comes from
vector meson ($\rho,\omega, K^*$ and $\phi$) exchanges. Some
authors \cite{Stancu,Shimisu} also studied short-range NN
repulsion as stemming from the Goldstone boson (and $\rho$-like)
exchanges on the quark level. It has been shown that these
interactions can substitute traditional OGE mechanism. Naturally,
we would like to ask which is the right mechanism for describing
the short range interactions? Is it from OGE or from vector meson
exchange or both important? This is interesting and challenging.
To study this problem, we propose the extended chiral SU(3) quark
model\cite{Dai} in which the vector meson exchanges are involved.
\section{The Extended Chiral $SU(3)$ Quark Model}
\subsection{The Model}
By generalizing the interaction Lagrangian, on the basis of the
chiral $SU(3)$ quark model \cite{Zhang}, we further added the term
of coupling between quark and vector meson field,
\begin{equation}
{\cal{L}}_{I}^{V} = -i g_{chv}\overline{\psi}\gamma_{\mu}
\vec{\varphi}_{\mu} \cdot \vec{\tau} \psi -i\frac{f_{chv}}{2
M_p}\overline{\psi}\sigma_{\mu \nu} \partial_{\nu}
\vec{\varphi}_{\mu} \cdot \vec{\tau} \psi. \label{diseqn}
\end{equation}
Therefore, we got the extended chiral $SU(3)$ quark model
\cite{Dai}, the Hamiltonian of the system in this extended model
can be written as
\begin{eqnarray}
& H & =\sum\limits_{i}T_i-T_G+\sum\limits_{i<j}V_{ij},
\label{diseqn}
\end{eqnarray}
\begin{eqnarray}
V_{ij} =V_{ij}^{Conf}+V_{ij}^{OGE}+V_{ij}^{Ch}, \label{diseqn}
\end{eqnarray}
\begin{eqnarray}
& V_{ij}^{Ch} & = \sum^{8}_{a=0} V^{s_a}_{ij} + \sum^{8}_{a=0}
    V^{ps_a}_{ij}+ \sum^{8}_{a=0} V^{v_a}_{ij}~.\label{diseqn}
\end{eqnarray}
Compared with the chiral $SU(3)$ quark model\cite{Zhang}, we have
 additional term  of quark vector field coupling potential $V^{v_a}_{ij}$
 which is derived from Lagrangian in the  Eq.~(1).

\subsection{Determine the Parameters}
   We have two initial input parameters: the harmonic-oscillator width parameter
$b_{u}$, and the up (down) quark mass. The coupling constant
$g_{ch}$ for scalar and pseudo-scalar chiral field coupling is
fixed by experimental value. For vector chiral field coupling, the
$g_{chv}$ and $f_{chv}$ are the coupling constants for vector
coupling and tensor coupling respectively. In the study of nucleon
resonance transition coupling to vector meson, Riska et al took
the value of  $g_{chv}=3.0$ and neglected the tensor coupling part
\cite{Riska}. From one boson exchange theory on baryon level, we
can obtain the two values between quark and baryon levels. All
meson masses, $m_{ps}$, $m_{s}$, and $m_{v}$, are taken to be the
experimental values, except for $\sigma$ meson, its mass is
treated as an adjustable parameter. The cut-off mass $\Lambda$,
indicating the chiral symmetry breaking scale is taken to be
$\Lambda^{-1}=0.18fm$. In the calculation, $\eta$ and $\eta'$
mesons are mixed by $\eta_1$ and $\eta_8$ with the mixing angle
$\theta_{\eta}=-23^{o}$. $\omega$ and $\phi$ are mixed by
$\omega_1$ and $\omega_8$ with the mixing angle
$\theta_{\omega}=-54.7^{o}$.
 After the parameters of chiral fields
are fixed, the one gluon exchange coupling constants $g_{u}$ can
be determined by the mass splits between N and  $\Delta$. The
confinement strength is fixed by the stability conditions of N.
The resultant model parameters are tabulated in Table 1.
\begin{table}[h]
\tbl{Model parameters and the binding energies of deuteron.}
{\begin{tabular}{@{}cccc@{}} \toprule &  chiral $SU(3)$ quark
model  & Extended  chiral $SU(3)$ quark model \\
 & & set I ~~~~~~~~~~~~set II  \\
\colrule
$b_u (fm)$             & 0.5      & 0.45  ~~~~~~~~~  0.45      \\
$g_{nn\pi}$            & 13.67    & 13.67  ~~~~~~~~~   13.67     \\
$g_{ch}$               & 2.621    & 2.621   ~~~~~~~~~  2.621     \\
$g_{chv}$              & 0        & 2.351   ~~~~~~~~~   1.972     \\
$f_{chv}/g_{chv}$      & 0        & 0      ~~~~~~~~~   2/3       \\
$m_{\sigma}(MeV)$      & 595      & 535    ~~~~~~~~~  547       \\
$g_u$                  & 0.886    & 0.293   ~~~~~~~~~  0.399     \\
$\alpha_s (g_u^2)$     & 0.785    & 0.086   ~~~~~~~~~   0.159     \\
$a_{uu}(MeV/fm^2)$     & 48.1     & 48.0     ~~~~~~~~~   42.9      \\
$B_{deu}(MeV)$         & 2.13     & 2.19    ~~~~~~~~~   2.14      \\
\botrule
\end{tabular}}
\end{table}
\section{Results and discussions}
The N-N phase shifts are calculated by solving a Resonating Group
Method (RGM) equation in the extended chiral $SU(3)$ quark model.
For comparison with results from different models, the results of
chiral $SU(3)$ quark model without vector meson exchanges
\cite{Zhang} are drawn with short-dashed curves, while long dashed
and solid curves are those obtained from the extended chiral
$SU(3)$ quark model \cite{Dai} with two different sets of
parameters (set I and set II). All S,P,D and F partial waves
calculated are very similar in both models, and consistent with
the experimental results \cite{Dai} . Here we just discuss two
main results:

\begin{romanlist}[(ii)]
\item   \textbf{The phase shifts of  S-wave}\\
 The results ( in Fig.1) show that the
$^1S_0$ phase shifts  are obviously improved in the extended
chiral $SU(3)$ quark model. The $^3S_1$ phase shifts of different
models are almost the same. To get the right trend of the phase
shift versus scattering energy, the size parameter $b_u$ is taken
with different values for these two models. $b_u = 0.50 fm$ in the
chiral $SU(3)$ quark model, and $b_u = 0.45 fm$ in the extended
chiral $SU(3)$ quark model. This means that the bare radius of
baryon becomes smaller when more meson clouds are included in the
model. This physical picture looks reasonable.

\item \textbf{The coupling constant of OGE }\\
When the vector meson field coupling is considered, the coupling
constant of OGE is greatly reduced by fitting the mass difference
between $\Delta$ and $N$. For both set I and II, $\alpha_s< 0.2$,
which is much smaller than the value $(0.785)$ of chiral $SU(3)$
quark model (see Table 1). It means that  the OGE interaction is
rather weak in the extended chiral $SU(3)$ quark model. Instead,
the vector meson exchanges play an important role for the short
range interaction between two quarks. Hence, mechanisms of the
quark-quark short range interactions of these two models are
totally different. In the chiral $SU(3)$ quark model, the short
range interaction is dominantly from OGE, and in the extended
chiral $SU(3)$ quark model, it is dominantly from vector meson
exchanges.
\end{romanlist}

\begin{figure}
\centerline{\psfig{file=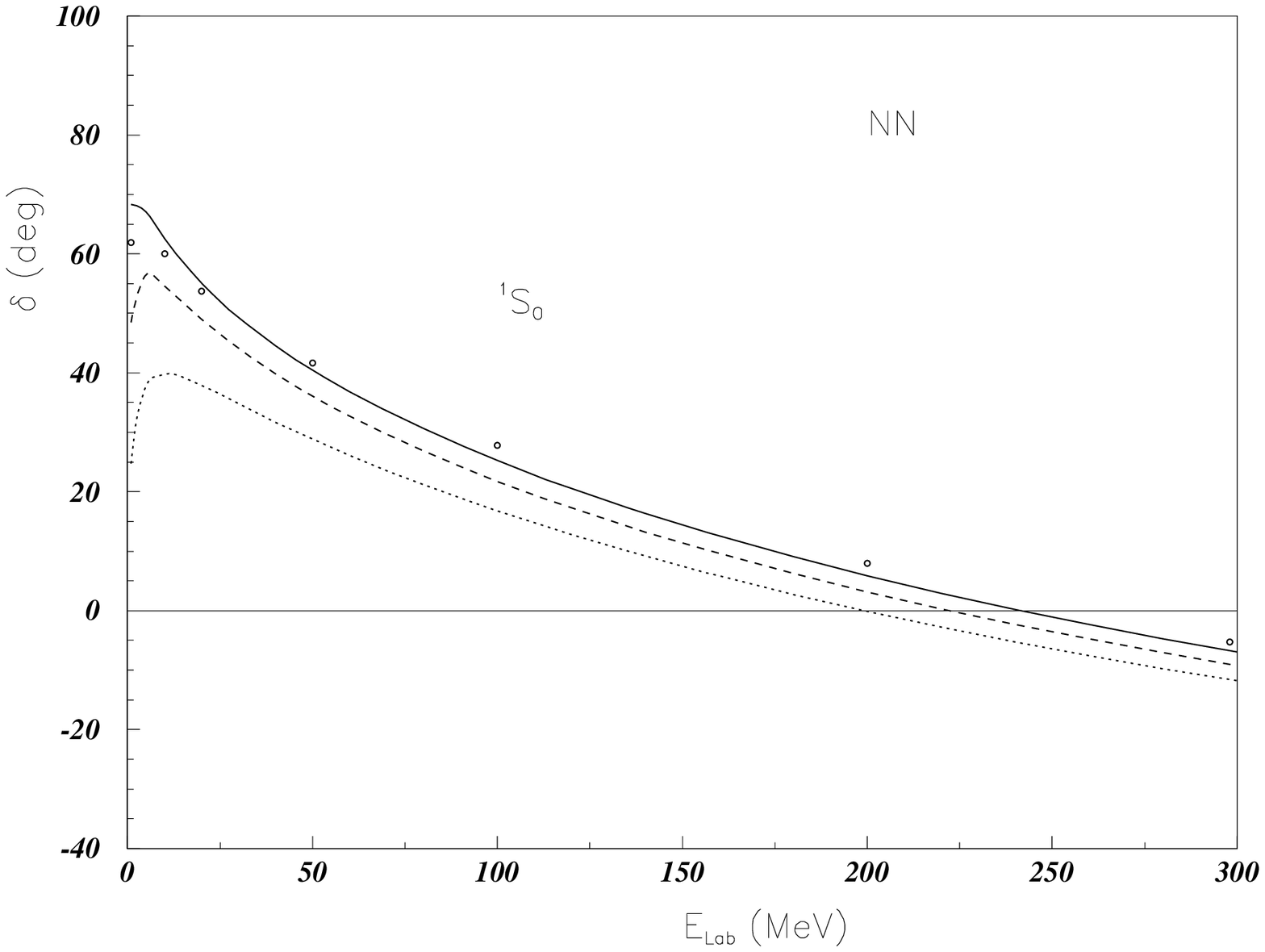,width=5cm}\psfig{file=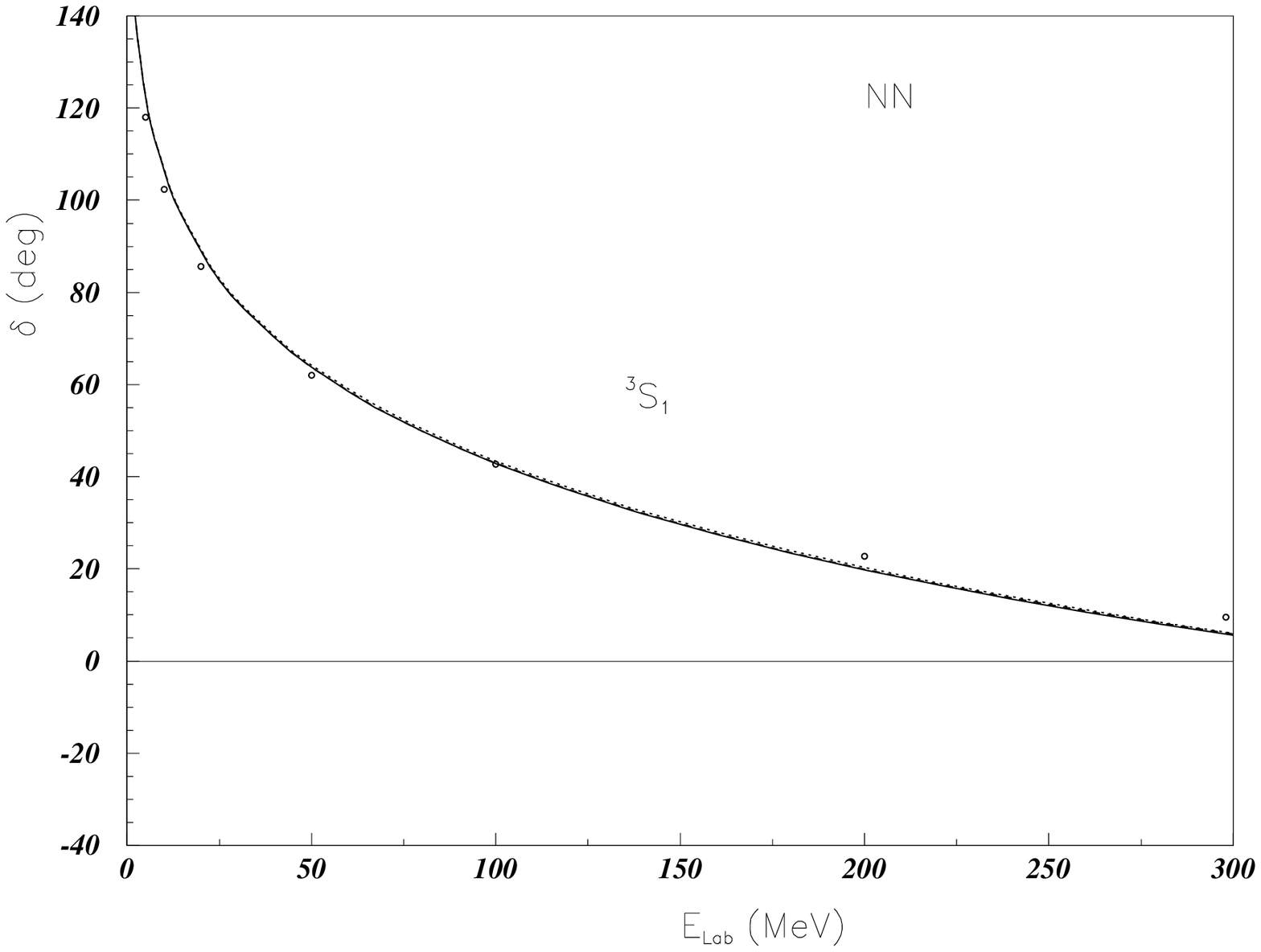,width=5cm}}
\vspace*{8pt} \caption{The NN scattering S-Wave phase shifts}
\end{figure}

 These features are reasonable and helpful in better understanding
the short range mechanism of the quark-quark interactions.

\end{document}